\documentclass[12pt]{article}
\usepackage{theorem}
\usepackage{amssymb,amsmath,mathrsfs, eucal}
\usepackage{amssymb}
\usepackage{graphicx}
\newtheorem{prop}{}[section]

{\theorembodyfont{\upshape} }
\newcommand{\boma}[1]{{\mbox{\boldmath $#1$} }}

\begin{document}
\def\Lip{{\mathcal A}}
\def\ellu{\ell_{*}}
\def\Sgr{\Gamma_{\rho}}
\def\aiz{a^i_{(0)}}
\def\aizn{a^i_{(0) \, n}}
\def\amuz{a^{\mu}_{(0)}}
\def\var{x}
\def\Zac{H}
\def\Zsinupa{\sin Y}
\def\Zsinupb{\sin(2 Y)}
\def\Zsinupc{\sin(3 Y)}
\def\Zsinupd{\sin(4 Y)}
\def\Zsinupe{\sin(5 Y)}
\def\Zsipama{\sin(\te - 5 Y)}
\def\Zsipamb{\sin(\te - 4 Y)}
\def\Zsipamc{\sin(\te - 3 Y)}
\def\Zsipamd{\sin(\te - 2 Y)}
\def\Zsipame{\sin(\te - Y)}
\def\Zsipanu{\sin\te}
\def\Zsipapa{\sin(\te + Y)}
\def\Sepapa{\sin(\te + Y)}
\def\Zsipapb{\sin(\te + 2 Y)}
\def\Sepapb{\sin(\te + 2 Y)}
\def\Zsipapc{\sin(\te + 3 Y)}
\def\Sepapc{\sin(\te + 3 Y)}
\def\Zsipbma{\sin(2 \te - 5 Y)}
\def\Zsipbmb{\sin(2 \te - 4 Y)}
\def\Zsipbmc{\sin(2 \te - 3 Y)}
\def\Zsipbmd{\sin(2 \te - 2 Y)}
\def\Zsipbme{\sin(2 \te - Y)}
\def\Zsipbnu{\sin (2 \te)}
\def\Zsipbpa{\sin(2 \te + Y)}
\def\Zsipbpb{\sin(2 \te + 2 Y)}
\def\Zsipbpc{\sin(2 \te + 3 Y)}
\def\Zsipcma{\sin(3 \te - 5 Y)}
\def\Zsipcmb{\sin(3 \te - 4 Y)}
\def\Zsipcmc{\sin(3 \te - 3 Y)}
\def\Zsipcmd{\sin(3 \te - 2 Y)}
\def\Zsipcme{\sin(3 \te - Y)}
\def\Zsipcnu{\sin (3 \te)}
\def\Zsipcpa{\sin(3 \te + Y)}
\def\Zsipcpb{\sin(3 \te + 2 Y)}
\def\Zsipcpc{\sin(3 \te + 3 Y)}
\def\Zsipdma{\sin(4 \te - 5 Y)}
\def\Zsipdmb{\sin(4 \te - 4 Y)}
\def\Zsipdmc{\sin(4 \te - 3 Y)}
\def\Zsipdmd{\sin(4 \te - 2 Y)}
\def\Zsipdme{\sin(4 \te - Y)}
\def\Zsipdnu{\sin (4 \te)}
\def\Zsipdpa{\sin(4 \te + Y)}
\def\Zsipdpb{\sin(4 \te + 2 Y)}
\def\Zsipdpc{\sin(4 \te + 3 Y)}
\def\Zsipema{\sin(5 \te - 5 Y)}
\def\Zsipemb{\sin(5 \te - 4 Y)}
\def\Zsipemc{\sin(5 \te - 3 Y)}
\def\Zsipemd{\sin(5 \te - 2 Y)}
\def\Zsipeme{\sin(5 \te - Y)}
\def\Zsipenu{\sin (5 \te)}
\def\Zsipepa{\sin(5 \te + Y)}
\def\Zsipepb{\sin(5 \te + 2 Y)}
\def\Zsipepc{\sin(5 \te + 3 Y)}
\def\Zsiqa{\sin(6 \te - 2 Y)}
\def\Zsiqb{\sin(6 \te)}
\def\Zsiqc{\sin(7 \te - 3 Y)}
\def\Zsiqd{\sin(7 \te - Y)}
\def\Zsiqe{\sin(8 \te - 2 Y)}
\def\Zsiqf{\sin(9 \te - 3 Y)}
\def\Zconupa{\cos Y}
\def\Zconupb{\cos(2 Y)}
\def\Zconupc{\cos(3 Y)}
\def\Zconupd{\cos(4 Y)}
\def\Zconupe{\cos(5 Y)}
\def\Zcopama{\cos(\te - 5 Y)}
\def\Zcopamb{\cos(\te - 4 Y)}
\def\Zcopamc{\cos(\te - 3 Y)}
\def\Zcopamd{\cos(\te - 2 Y)}
\def\Zcopame{\cos(\te - Y)}
\def\Zcopanu{\cos\te}
\def\Zcopapa{\cos(\te + Y)}
\def\Zcopapb{\cos(\te + 2 Y)}
\def\Zcopapc{\cos(\te + 3 Y)}
\def\Zcopbma{\cos(2 \te - 5 Y)}
\def\Zcopbmb{\cos(2 \te - 4 Y)}
\def\Zcopbmc{\cos(2 \te - 3 Y)}
\def\Zcopbmd{\cos(2 \te - 2 Y)}
\def\Zcopbme{\cos(2 \te - Y)}
\def\Zcopbnu{\cos (2 \te)}
\def\Zcopbpa{\cos(2 \te + Y)}
\def\Zcopbpb{\cos(2 \te + 2 Y)}
\def\Zcopbpc{\cos(2 \te + 3 Y)}
\def\Zcopcma{\cos(3 \te - 5 Y)}
\def\Zcopcmb{\cos(3 \te - 4 Y)}
\def\Zcopcmc{\cos(3 \te - 3 Y)}
\def\Zcopcmd{\cos(3 \te - 2 Y)}
\def\Zcopcme{\cos(3 \te - Y)}
\def\Zcopcnu{\cos (3 \te)}
\def\Zcopcpa{\cos(3 \te + Y)}
\def\Zcopcpb{\cos(3 \te + 2 Y)}
\def\Zcopcpc{\cos(3 \te + 3 Y)}
\def\Zcopdma{\cos(4 \te - 5 Y)}
\def\Zcopdmb{\cos(4 \te - 4 Y)}
\def\Zcopdmc{\cos(4 \te - 3 Y)}
\def\Zcopdmd{\cos(4 \te - 2 Y)}
\def\Zcopdme{\cos(4 \te - Y)}
\def\Zcopdnu{\cos (4 \te)}
\def\Zcopdpa{\cos(4 \te + Y)}
\def\Zcopdpb{\cos(4 \te + 2 Y)}
\def\Zcopdpc{\cos(4 \te + 3 Y)}
\def\Zcopema{\cos(5 \te - 5 Y)}
\def\Zcopemb{\cos(5 \te - 4 Y)}
\def\Zcopemc{\cos(5 \te - 3 Y)}
\def\Zcopemd{\cos(5 \te - 2 Y)}
\def\Zcopeme{\cos(5 \te - Y)}
\def\Zcopenu{\cos (5 \te)}
\def\Zcopepa{\cos(5 \te + Y)}
\def\Zcopepb{\cos(5 \te + 2 Y)}
\def\Zcopepc{\cos(5 \te + 3 Y)}
\def\x{{\scriptscriptstyle{X}}}
\def\y{{\scriptscriptstyle{Y}}}
\def\h{{\scriptscriptstyle{H}}}
\def\r{{\tt r}}
\def\ti{{\mathfrak t}}
\def\QQ{\mathscr D}
\def\sc{{\scriptscriptstyle{\bullet}}}
\def\k{\boma{k}}
\def\PP{{\mathscr P}}
\def\M{A}
\def\bomu{\boma{\mu}}
\def\bonu{\boma{\nu}}
\def\bosi{\boma{\sigma}}
\def\S{S}
\def\dist{\mbox{dist}}
\def\XX{\Xi}
\def\YY{\mbox{{\rm H}}}
\def\xx{\xi}
\def\yy{\eta}
\def\m{m}
\def\Np{\mathfrak N}
\def\Lp{\mathfrak L}
\def\EN{\mathcal N}
\def\EL{\mathcal L}
\def\Ti{\mathscr T}
\def\Tim{T_1}
\def\kk{K}
\def\K{{\tt K}}
\def\R{{\tt R}}
\def\RR{\mathscr{R}}
\def\Te{\Theta}
\def\ttet{\dot \Theta}
\def\A{{\tt A}}
\def\R{{\tt R}}
\def\L{{\tt L}}
\def\J{{\tt J}}
\def\I{{\tt I}}
\def\H{{\tt H}}
\def\X{{\tt X}}
\def\Y{{\tt Y}}
\def\Tt{{\tt T}}
\def\vel{{\tt v}}
\def\mm{\sigma}
\def\aa{\alpha_{0}}
\def\aad{\alpha_{\delta}}
\def\aadp{\alpha_{\delta'}}
\def\sca{{\scriptstyle{\,\bullet\,}}}
\def\pp{{\mathfrak p}}
\def\qq{{\mathfrak q}}
\def\rr{{\mathfrak r}}
\def\ha{I^1}
\def\ka{I^2}
\def\lau{\lambda_1}
\def\lad{\lambda_2}
\def\tr{\mbox{{\rm tr}}}
\def\v{v}
\def\p{p}
\def\q{q}
\def\u{u}
\def\w{w}
\def\z{z}
\def\Gi{\Lambda}
\def\Gam{\Lambda_{\dag}}
\def\MM{\mathscr{M}}
\def\UU{\mathscr{U}}
\def\FF{\dd{\partial \overline{f} \over \partial I}}
\def\GG{\mathscr{G}}
\def\SS{\mathscr{S}}
\def\RR{\mathscr{R}}
\def\NN{\mathscr{N}}
\def\AA{\mathscr{A}}
\def\BB{\mathscr{B}}
\def\CC{\mathscr{C}}
\def\DD{\mathscr{D}}
\def\EE{\mathscr{E}}
\def\LL{\mathscr{L}}
\def\HH{\mathscr{H}}
\def\Ten{\mbox{T}}
\def\Tuu{\Ten^1_1(\reali^{d})}
\def\Tdz{\Ten^2_0(\reali^{d})}
\def\Tud{\Ten^1_2(\reali^d)}
\def\uno{1_d}
\def\ppsi{\vartheta}
\def\PPsi{\Theta}
\def\N{V}
\def\half{{1 \over 2}}
\def\II{M}
\def\JJ{N}
\def\ga{\gamma_{\scriptscriptstyle{E}}}
\def\mmu{\nu}
\def\nnu{\mu}
\def\f{g}
\def\Fs{\mathscr{G}}
\def\Kap{\mathscr{K}}
\def\STB{\scriptsize B}
\def\STBB{\scriptsize BB}
\def\STF{\scriptsize F}
\def\STFF{\scriptsize FF}
\def\TB{\scriptsize(B)}
\def\TBB{\scriptsize(BB)}
\def\TF{\scriptsize(F)}
\def\TFF{\scriptsize(FF)}
\def\XXX{\mathscr X}
\def\Piu{\mathscr P}
\def\Men{\mathscr N}
\def\ffi{\varphi}
\def\ES{{\mathcal S}}
\def\KK{{\mathscr K}}
\def\KKp{{\mathscr K}'}
\def\TT{{\mathfrak T}}
\def\kp{k'}
\def\scrscr{\scriptscriptstyle}
\def\scr{\scriptstyle}
\def\dd{\displaystyle}
\def\B{ B_{\mbox{\scriptsize{\textbf{C}}}} }
\def\Bc{ \overline{B}_{\mbox{\scriptsize{\textbf{C}}}} }
\def\ppartial{\overline{\partial}}
\def\d{d}
\def\e{e}
\def\Hinf{ H^{\infty}(\reali^d, \complessi) }
\def\Hn{ H^{n}(\reali^d, \complessi) }
\def\Hm{ H^{m}(\reali^d, \complessi) }
\def\Ha{ H^{\d}(\reali^d, \complessi) }
\def\Ld{L^{2}(\reali^d, \complessi)}
\def\Lpi{L^{p}(\reali^d, \complessi)}
\def\Lq{L^{q}(\reali^d, \complessi)}
\def\Lr{L^{r}(\reali^d, \complessi)}
\def\Knb{K^{best}_n}
\def\D{\mbox{{\tt D}}}
\def\g{ {\textbf g} }
\def\QQQ{ {\textbf Q} }
\def\AAA{ {\textbf A} }
\def\gr{\mbox{graph}~}
\def\Q{$\mbox{Q}_a$~}
\def\PZ{$\mbox{P}^{0}_a$~}
\def\PZAL{$\mbox{P}^{0}_\alpha$~}
\def\PL{$\mbox{P}^{1/2}_a$~}
\def\PU{$\mbox{P}^{1}_a$~}
\def\PK{$\mbox{P}^{k}_a$~}
\def\PKU{$\mbox{P}^{k+1}_a$~}
\def\PI{$\mbox{P}^{i}_a$~}
\def\Pell{$\mbox{P}^{\ell}_a$~}
\def\PTM{$\mbox{P}^{3/2}_a$~}
\def\AZ{$\mbox{A}^{0}_r$~}
\def\AU{$\mbox{A}^{1}$~}
\def\epsilona{\epsilon^{\scriptscriptstyle{<}}}
\def\epsilonb{\epsilon^{\scriptscriptstyle{>}}}
\def\lgraffa{ \mbox{\Large $\{$ } \hskip -0.2cm}
\def\rgraffa{ \mbox{\Large $\}$ } }
\def\restriction{ \stackrel{\setminus}{~}\!\!\!\!|~}
\def\m{m}
\def\Fre{Fr\'echet~}
\def\ap{{\scriptscriptstyle{ap}}}
\def\fiap{\varphi_{\ap}}
\def\BBB{ {\textbf B} }
\def\EEE{ {\textbf E} }
\def\FFF{ {\textbf F} }
\def\TTT{ {\textbf T} }
\def\KKK{ {\textbf K} }
\def\FFi{ {\bf \Phi} }
\def\a{a}
\def\ep{\varepsilon}
\def\parn{\par}
\def\teta{M}
\def\elle{L}
\def\ro{\rho}
\def\al{\alpha}
\def\si{\sigma}
\def\vsi{\varsigma}
\def\kap{\kappa}
\def\be{\beta}
\def\de{\delta}
\def\la{{\mathfrak l}}
\def\mi{{\mathfrak v}}
\def\en{{\mathfrak n}}
\def\em{{\mathfrak m}}
\def\te{\vartheta}
\def\tet{\dot \vartheta}
\def\It{\dot I}
\def\Jta{J'}
\def\om{\omega}
\def\ch{\chi}
\def\complessi{{\textbf C}}
\def\reali{{\textbf R}}
\def\interi{{\textbf Z}}
\def\naturali{{\textbf N}}
\def\bT{{\textbf T}}
\def\T1{{\textbf T}^{1}}
\def\Jp{{\hat{J}}}
\def\Pp{{\hat{P}}}
\def\Pip{{\hat{\Pi}}}
\def\Vp{{\hat{V}}}
\def\Ep{{\hat{E}}}
\def\Fp{{\hat{F}}}
\def\Gp{{\hat{G}}}
\def\Ip{{\hat{I}}}
\def\Tp{{\hat{T}}}
\def\Mp{{\hat{M}}}
\def\La{\Lambda}
\def\Upsi{\Upsilon}
\def\Lap{{\hat{\Lambda}}}
\def\Sip{{\hat{\Sigma}}}
\def\Upsig{{\check{\Upsilon}}}
\def\Kg{{\check{K}}}
\def\ellp{{\hat{\ell}}}
\def\j{j}
\def\jp{{\hat{j}}}
\def\cir{{\scriptscriptstyle \circ}}
\def\circa{\thickapprox}
\def\vain{\rightarrow}
\def\leqs{\leqslant}
\def\geqs{\geqslant}
\def\ss{s}
\def\vains{\stackrel{\ss}{\rightarrow}}
\def\parnn{\par \noindent}
\def\salto{\vskip 0.2truecm \noindent}
\def\spazio{\vskip 0.5truecm \noindent}
\def\vs1{\vskip 1cm \noindent}
\def\fine{$\Box$}
\newcommand{\rref}[1]{(\ref{#1})}
\def\beq{\begin{equation}}
\def\feq{\end{equation}}
\def\beqq{\begin{eqnarray}}
\def\feqq{\end{eqnarray}}
\def\barray{\begin{array}}
\def\farray{\end{array}}
\makeatletter \@addtoreset{equation}{section}
\renewcommand{\theequation}{\thesection.\arabic{equation}}
\makeatother
\begin{titlepage}
\begin{center}
{\huge On the averaging principle for \\ one-frequency systems. \\
Seminorm estimates for the error}
\end{center}
\vspace{1truecm}
\begin{center}
{\large
Carlo Morosi${}^1$, Livio Pizzocchero${}^2$} \\
\vspace{0.5truecm} ${}^1$ Dipartimento di Matematica, Politecnico
di
Milano, \\ P.za L. da Vinci 32, I-20133 Milano, Italy \\
e--mail: carmor@mate.polimi.it \\
${}^2$ Dipartimento di Matematica, Universit\`a di Milano\\
Via C. Saldini 50, I-20133 Milano, Italy\\
and Istituto Nazionale di Fisica Nucleare, Sezione di Milano, Italy \\
e--mail: livio.pizzocchero@mat.unimi.it
\end{center}
\vspace{1truecm}
\begin{abstract}
We extend some previous results of our work \cite{uno} on the error of the averaging method, in the
one-frequency case.
The new error estimates apply to any separating
family of seminorms on the space of the actions; they generalize our previous estimates
in terms of the Euclidean norm. For example, one can use the new approach to get
separate error estimates for each action coordinate.
An application to rigid body under damping is presented. In a
companion paper \cite{tre}, the same method will be applied to the motion
of a satellite around an oblate planet.
\end{abstract}
\vspace{1truecm} \noindent \textbf{Keywords:} Slow and fast motions,
perturbations, averaging method.
\par \vspace{0.4truecm} \noindent \textbf{AMS 2000 Subject
classification}: 70K65, 70K70, 34C29, 70H09, 37J40.
\end{titlepage}
\section{Introduction.}
\label{intro}
It is often stated by applied mathematicians that a good theorem on differential equations
is one outlining a computational method for their solutions; if the method is approximate,
quantitative error estimates should be provided. \parn
In the case of ODEs with slow variables ("actions") and fast angular variables,
averaging over the angles is a well-known approximation technique; in the literature,
the error of this method has been discussed mainly from a qualitative viewpoint, even
in the simple case of one frequency (i.e., one angle only). The classical, qualitative estimates for
this case (see e.g. \cite{Arn}) have the form
\beq \I(t) - \J(\ep t) = O(\ep) \qquad \mbox{for $t \in [0,O(1/\ep))$} \label{qual} \feq
(uniformly in $t$) for a perturbation proportional to a parameter $\ep$,
in the limit $\ep \vain 0^{+}$; here, $\I(t)$ are the actions at time $t$, and
$\J(\ep t)$ is their approximation obtained from averaging (see paragraph 1A for more details).
\parn
In a previous paper \cite{uno}, we have proposed in place of
\rref{qual} a \textsl{fully quantitative} error estimate for the
one-frequency averaging; this has the form
\beq |\I(t) - \J(\ep t)| \leqs \ep \en(\ep t) \qquad \mbox{for $t \in [0,U/\ep)$}~,
\label{form1} \feq where $|~|$ is the Euclidean norm on the space
of the actions and $\en$ is a computable function, determined by an
integral inequality; $U$ is a
specified nonnegative constant, defining quantitatively the time
interval where the estimate holds. (In fact, in some special cases
considered in \cite{uno} the estimate holds even for very large
values of $U$, e.g., $U \simeq 1/\ep$).
Let us repeat here a comment already done in the cited work: the idea of a really quantitative
approach to the averaging methods has attracted little attention up to now,
a notable exception being \cite{Smi} that, in \cite{uno}, we have briefly compared with
our approach. \parn
The present work is an improvement of \cite{uno} proposing more detailed error
estimates, e.g., a separate bound on each component of the
actions. These componentwise bounds are seen as a special case of
a more general framework, where the estimates are expressed in
terms of any separating family of seminorms on the space of the
actions (a notion to be defined in the sequel). Our general
estimates will take the form \beq |\I(t) - \J(\ep t)|^\mu \leqs
\ep \en^\mu(\ep t) \qquad \mbox{for $\mu \in M$, $t \in
[0,U/\ep)$}~, \label{form2} \feq where $(|~|^\mu)_{\mu \in M}$ is
the family of seminorms, labeled by a (finite) index set $M$.
\parn
To show the effectiveness of these bounds, in the present
work we give a simple example related to rigid body dynamics. A more engaging
application, concerning the motion of a satellite around an oblate planet, will
be presented in the companion paper \cite{tre}.
\parn
The forthcoming paragraphs 1A-1D introduce the following topics: the setting of \cite{uno}
for one-frequency averaging, that we use partly in this paper; the new error estimates
developed in the present work; the motivations to consider these refinements,
and to formulate them in the language of seminorms; the organization of the paper.
\salto
\textbf{1A. One-frequency averaging, in the framework of \cite{uno}.}
We consider an open set $\Lambda$ of $\reali^d$ and the one-dimensional torus $\bT$
(referred to as the spaces of the actions and of the angular variable):
\beq \Lambda = \{ I = (I^i)_{i=1,...,d} \} \subset \reali^d~, \qquad \bT :=
\reali/ 2 \pi \interi = \{ \te \}~. \feq
We suppose to be given a one-frequency system with a perturbation
$\ep f$ on the actions and $\ep g$ on the angle: more precisely,
we have a Cauchy problem
\beq \left\{\barray{ll} d \I/dt = \ep f(\I,\Te)~, &\quad \I(0) = {I_0}~, \\
d \Te/ d t = \om(\I) +\ep g(\I,\Te)~, &\quad \Te(0) = {\te_0}~, \farray
\right. \label{pert} \feq
under the assumptions
\beq f = (f^i)_{i=1,...,d} \in C^\m(\Gi \times \bT, \reali^d),~
g \in C^\m(\Gi \times \bT, \reali),~\om \in C^\m(\Gi, \reali)~(m \geqs 2)~, \feq
$$ \om(I) \neq 0 \mbox{ for all $I \in \Gi$}~, \quad
I_0 \in \Lambda, \te_0 \in \bT~, \qquad \ep > 0~; $$
the maximal solution (in the future) of \rref{pert} is a $C^{m+1}$ function
\beq (\I, \Te) : [0,T) \vain \Lambda \times \bT~, \qquad t \mapsto (\I(t),
\Te(t))~. \feq
Throughout the paper, the initial data $I_0, \te_0$ and the perturbation parameter $\ep$ are
fixed; for this reason, we do not indicate
the dependence  of $(\I,\Te)$ and other functions on these objects. Needless to say,
we are mainly interested in the case of small $\ep$. \parn
The averaged system associated to \rref{pert} is
\beq {d \J \over d \tau} = \overline{f}(\J)~, \qquad  \J(0) = I_0~,
\label{av} \feq
$$ \overline{f} = (\overline{f^i})_{i=1,...,d} \in C^\m(\Gi,\reali^d)~,
\qquad I \mapsto \overline{f}(I) :=
{1 \over 2 \pi} \int_{\bT} d \te~f(I,\te)~; $$
the maximal solution (in the future) is a $C^{m+1}$
function
\beq \J : [0,W) \mapsto \Lambda~, \qquad \tau \mapsto \J(\tau)~.
\feq
The error of the averaging method is the function $t \mapsto \I(t) - \J(\ep t)$
(defined whenever $\I(t)$ and $\J(\ep t)$ exist); equivalently, one can consider the function
\beq \L : t \mapsto \L(t) := {1 \over \ep} [\I(t) - \J(\ep t)]~. \feq
In \cite{uno} we have
put the attention on the Euclidean norm
\beq | \L(t) | = \sqrt{\sum_{i=1}^d \L^i(t)^2}~; \label{euclidea} \feq
under natural conditions, we have derived for it a quantitative estimate
\beq | \L(t) | \leqs \en(\ep t) \qquad \mbox{for $t \in [0,U/\ep)$} \label{togen} \feq
(which is the same as \rref{form1}), where $\en : [0,U) \vain [0,+\infty)$ is
a function determined by a fully explicit algorithm. To compute $\en$, one must solve
an integral inequality or a related differential equation on $[0,U)$, a task that
in typical cases is performed numerically; however, for $\ep$ small this operation is
much faster than the direct numerical solution of the perturbed system \rref{pert}
for $t$ in the long interval $[0,U/\ep)$.
\salto
\textbf{1B. Some variants in analyzing $\boma{\L(t)}$.} In view of
applications, the following variants can be of interest: \parnn
(a) estimating a norm of $\L(t)$ different from \rref{euclidea}; \parnn
(b) giving separate estimates on the absolute values $|\L^i(t)|$ of the
components; \parnn
(c) considering a partition $\PP = \{\S,\S'...\}$ of $\{1,...,d\}$ into
(nonempty) subsets $\S, \S',...$ and
estimating the components of $\L(t)$ in each subset: for example,
one could analyze the quantities
\beq \sqrt{\sum_{i \in \S} \L^i(t)^2}~,~\sqrt{\sum_{i \in \S'}
\L^i(t)^2}~,...~. \feq
Here are some reasons to study each component separately, or to
group them into subsets: the components could measure physically nonhomogeneous
quantities; one expects relevant differences in their numerical values,
even in the orders of magnitude. All these facts will occur in the example
of Section \ref{esrigi}, related to rigid body dynamics. \parn
\salto
\textbf{1C. General estimates for $\boma{\L(t)}$ via seminorms.}
A unified way to treat (a) (b) (c) and other situations is to
consider on $\reali^d$
a \textsl{separating family of seminorms}, and use them to estimate
$\L(t)$. Let us recall that a seminorm on $\reali^d$ is a map
\beq \reali^d \vain [0,+\infty)~, \qquad X \mapsto | X |~, \feq
homogeneous and subadditive:
\beq | \lambda X | = | \lambda | | X |~, \qquad | X + Y | \leqs | X | + | Y |
\qquad \mbox{for $X, Y \in \reali^d$, $\lambda \in \reali$} \label{homog} \feq
($|\lambda|$ is the absolute value of $\lambda$; the first relation, with
$\lambda = 0$,  gives $| 0 | = 0$).
An example of a seminorm is the function $|~ |^i$ on $\reali^d$,
where $i$ is any integer in $\{1,...,d\}$ and
\beq | X |^i := | X^i | \label{casei} \feq
for all $X = (X^1,...,X^d) \in \reali^d$; more generally, if $S$ is a
(nonempty) subset of $\reali^d$ we can define
a seminorm $| ~|^{S}$ on $\reali^d$, setting
\beq | X |^S := \sqrt{ \sum_{i \in S} (X^i)^2}~. \label{cases} \feq
A \textsl{norm} on $\reali^d$ can be defined as a seminorm with the
supplementary \textsl{separation} property
\beq | X | = 0 ~~\Rightarrow~~ X = 0~. \feq
Clearly, this property is lacking (for $d>1$) in the example \rref{casei}; it is also
lacking in \rref{cases}, unless
$S = \{1,...,d\}$. However, these examples with variable $i$ or $S$, and
other situations, carry to \textsl{families} of seminorms possessing
the separation property in a collective sense. To be precise, a \textsl{separating family
of seminorms} on
$\reali^d$ is a family $(| ~|^{\mu})_{\mu \in M}$, where $M$ is a finite set,
such that $|~|^{\mu}$ is a seminorm for each $\mu$ and, for all $X \in
\reali^d$,
\beq | X |^{\mu} = 0 \quad \mbox{for each $\mu \in M$}~~ \Rightarrow~~ X =
0~. \feq
An example of a separating family is formed by all the seminorms \rref{casei},
with $i$ ranging in $\{1,...,d\}$. Another example is the family \rref{cases},
labeled by the subsets $S$ in a partition $\PP$ of $\{1,...,d\}$. \parn
Throughout the paper, our estimates for $\L(t)$ will concern the
nonnegative quantities
\beq | \L(t) |^{\mu} \qquad (\mu \in M) \feq
for any chosen separating family of seminorms on $\reali^d$. Case (a) of the
previous paragraph
corresponds to the choice $M = \{1\}$ and $|~|^1 =$ a norm $|~|$ on
$\reali^d$; case (b) corresponds to the family \rref{casei} with $M = \{1,..., d\}$, and
case (c) to the family \rref{cases} with $M = \PP$.
\salto
\textbf{1D. Organization of the paper.} Section \ref{semin} is the main body
of the paper: after recalling a basic Lemma from \cite{uno}, we construct the general framework
to estimate $\L$ through a separating family of seminorms.
The conclusion is a set of inequalities
\beq | \L(t) |^\mu \leqs \en^\mu(\ep t) \qquad \mbox{for $\mu \in M$,
$t \in [0,U/\ep)$}  \feq
(i.e., of the form \rref{form2}), where the estimators $\en^\mu : [0,U) \vain [0,+\infty)$
are determined solving a system of integral inequalities (Proposition \ref{mainprop}), or of differential
equations  related to them (Proposition \ref{proprinc}).
Section \ref{esrigi} presents an example, arising
from the dynamics of a rigid body under damping; this was introduced
in \cite{uno} and will be reconsidered from the present viewpoint,
deriving separate error estimates for each one of the two actions. The Appendices \ref{apmain}, \ref{apprinc}
contain the proofs of the previously mentioned Propositions. \parn
In spite of the frequent reference to \cite{uno}, in writing the present paper we have tried to make
it reasonably self-contained.
\vskip 0.4cm \noindent
\vfill \eject \noindent
\section{Main results.}
\label{semin}
\textbf{2A. Notations.} (i) Throughout the paper,
vectors of $\reali^d$ are written with upper indices: $X = (X^i)_{i=1,...,d}$,
as already done in the Introduction. Due to the fact that $\reali^d$ has a canonical
basis, the tensors on $\reali^d$ of any type $(p, q)$ can be identified with
tables of real numbers, that we write in the usual style with $p$ upper and $q$ lower indices.
In the sequel we will often use the tensor spaces
$$ \Tuu = \{ \AA = (\AA^{i}_{j})~|~\AA^{i}_{j} \in \reali~ \mbox{for
$i,j=1,...,d$}
\}~, $$
\beq \Tdz = \{ \BB = (\BB^{ij})~|~\BB^{ij} \in \reali~ \mbox{for $i,j=1,...,d$}
\}~, \feq
$$ \Tud = \{ \CC = (\CC^{i}_{j k})~|~\CC^{i}_{j k} \in \reali~ \mbox{for
$i,j,k=1,...,d$}
\}~. $$
We use systematically Einstein's summation convention on repeated upper and lower indices.
Let $X, Y \in \reali^d$, $\AA, \DD \in \Tuu$ and $\CC \in \Tud$; then,
$\AA X \in \reali^d$ and $\CC X Y \in \reali^d$ are the vectors of components
$(\AA X)^i = \AA^{i}_{j} X^{j}$, $(\CC X Y)^i =
\CC^{i}_{j k} X^{j} Y^{k}$; $\AA \DD, \CC X \in \Tuu$ are the tensors of components
$(\AA \DD)^{i}_{k} = \AA^{i}_{j} \DD^{j}_{k}$, $(\CC X)^{i}_{k} = \CC^{i}_{j k} X^{j}$ . \salto
(ii) We fix on $\reali^d$ a separating family of seminorms
\beq |~|^{\mu} \qquad (\mu \in M)~, \label{fama} \feq
with $M$ a finite set. To go on, we need some seminorm families on the
tensor spaces
$\Tuu$ and $\Tud$; of course a seminorm on $\Tuu$ is a homogeneous,
subadditive map
\beq | ~| : \Tuu \vain [0,+\infty)~, \qquad \AA \mapsto | \AA | \feq
and a seminorm on $\Tud$ is defined similarly. \parn
Keeping fixed the family \rref{fama}, a
\textsl{consistent family of seminorms} on $\Tuu$ is one of the form
\beq |~|^{\mu}_{\nu} \qquad (\mu,\nu \in M)~,\feq
with the property
\beq |\AA X|^{\mu} \leqs |\AA|^{\mu}_{\nu} \, |X|^{\nu} \qquad \mbox{for
all $\AA \in \Tuu$,
$X \in \reali^d$ and $\mu \in M$}~ \label{cons1} \feq
(here and in the sequel, Einstein's summation convention is also employed for
repeated indices with values in $M$).
Similarly, a \textsl{consistent family of seminorms} on $\Tud$ is a family of seminorms
\beq |~|^{\mu}_{\nu \kap} \qquad (\mu,\nu,\kap \in M)~,\feq
such that
\beq |\CC X Y|^{\mu} \leqs |\CC|^{\mu}_{\nu \kap} |X|^{\nu} |Y|^{\kap}
\qquad \mbox{for all $\CC \in \Tuu$,
$X,Y \in \reali^d$ and $\mu \in M$}~. \label{cons2} \feq
The existence of such consistent families can be proved using the separation
property of \rref{fama} ({\footnote{This follows from much more general results on
multilinear maps and seminorms that can be found, e.g., in \cite{Hor}.}}).
Of course, if we use on $\reali^d$ the seminorms
$|~|^i$ of Eq. \rref{casei} we have on $\Tuu$ and $\Tud$ the following consistent
families of seminorms, also taking the absolute values of the tensor components:
\beq | \AA |^{i}_{j} := | \AA^{i}_{j} |~, \qquad | \CC |^{i}_{j k} :=  | \CC^{i}_{j k} |
\qquad (i,j,k = 1,...,d)~.
\label{consof} \feq
(iii) In the sequel we intend
\beq \Gam := \{ (I,\delta I) \in \Gi \times \reali^d~|~[I, I + \delta I ]
\subset \Lambda \}~, \label{deg} \feq
where $[I, I + \delta I]$ is the closed segment in $\reali^d$ with the
indicated extremes.
\salto
\textbf{2B. The integral equation for $\boma{\L}$.} We consider the perturbed and averaged systems
\rref{pert} \rref{av}, for fixed
$\ep > 0$ and initial data $I_0, \te_0$. We introduce the functions
$s \in C^\m(\Gi \times \bT,\reali^d)$ and
$\p \in C^{\m-1}(\Gi \times \bT, \reali^d)$ such that
\beq f = \overline{f} + \omega~ {\partial s \over \partial \te}~, \quad
\overline{s} = 0~; \qquad
\p := {\partial s \over \partial I} f + {\partial s \over \partial \te} g~; \label{thef} \feq
these equations, and the forthcoming ones are always understood in the tensorial sense
({\footnote{For better clarity, let us give only some examples.
The equivalents in components of Eq. \rref{thef} are
$$ f^i = \overline{f^i} + \omega~ {\partial s^i \over \partial \te}~, \quad
\overline{s^i} = 0~; \qquad
\p^i := {\partial s^i \over \partial I^j} f^j + {\partial s^i \over \partial \te} g~. $$
In the forthcoming Eq.s \rref{em} and \rref{equel}, the relations about
for $\MM, \overline{f}$ and $\overline{f}$, $\HH$ mean, respectively:
$$\dd{\MM^{i}_{k} := {\partial^2 \overline{f^i} \over \partial I^j \partial I^k }\, \overline{f^j} -
{\partial \overline{f^i} \over \partial I^j} \, {\partial \overline{f^j} \over \partial I^k}}~; $$
$$ \overline{f}^i(I + \delta I) = \overline{f}^i(I) +
{\partial \overline{f}^i \over \partial I^j}(I) \delta I^j +
{1 \over 2} \HH^{i}_{j k}(I,\delta I) \delta I^j \delta I^k~. $$
}}). \parnn
From now on, $U$ stands for an element of $(0,+\infty]$.
\begin{prop}
\label{lemma1}
\textbf{Lemma.}
Suppose the solution $\J$ of \rref{av}
exists for $\tau \in [0,U)$.
Denote by $\R : [0,U) \vain \Tuu$, $\tau \mapsto \R(\tau)$ and
$\K :  [0,U) \vain \reali^d$, $\tau\mapsto \K(\tau)$
the solutions of
\beq {d \R \over d \tau} = \FF(\J) \,\R~,
\qquad \R(0) = \uno~; \label{sistr} \feq
\beq {d \K \over d \tau} = \FF(\J) \,\K + \overline{\p}(\J)~,
\qquad \K(0) = 0~ \label{sistk} \feq
(these exist and are $C^\m$; $\R(\tau)$ is an invertible matrix for all $\tau \in [0,U)$, and
$\K(\tau) = \R(\tau) \int_{0}^{\tau} d \tau' \, \R(\tau')^{-1} \overline{p}(\J(\tau'))$.
For $d=1$, $\R(\tau) = \exp \int_{0}^{\tau} d \tau' \, \FF(\J(\tau')) \in (0,+\infty)$).
\parn
Furthermore, assume that the solution $(\I,\Te)$ of the perturbed system
\rref{pert} exists for $t \in [0, U / \ep)$, with $(\J(\ep t), \I(t) - \J(\ep t)) \in \Gam$. Finally, define
\beq \L : [0,U/\ep) \vain \reali^d~, \qquad
t \mapsto \L(t) :=
{1 \over \ep} \,[ \I(t) - \J(\ep t) ]~. \label{defr} \feq
Then, for $t \in [0,U/\ep)$,
\beq  \L(t) = s(\I(t), \Te(t)) - \R(\ep t) \, s(I_0, \te_0) - \K(\ep t)  \label{inseq}\feq
$$ - \ep \Big(\w(\I(t), \Te(t)) - \FF(\J(\ep t)) \, \v(\I(t), \Te(t)) \Big)   $$
$$ + \,\ep^2 \R(\ep t) \int_{0}^t d t' \,
\R^{-1}(\ep t') \Big(\u(\I(t'),\Te(t')) -\FF(\J(\ep t')) (\w +\q)(\I(t'),\Te(t'))  $$
$$ - \MM(\J(\ep t')) \v(\I(t'),\Te(t')) - \GG(\J(\ep t'), \ep  \L(t')) \L(t')
+ {1 \over 2} \HH(\J(\ep t'), \ep \L(t')) \, \L(t')^2\Big)~. $$
In the above, $v \in C^\m(\Gi \times \bT,\reali^d)$,
$q, \w \in C^{\m-1}(\Gi \times \bT, \reali^d)$,
$\u \in C^{\m-2}(\Gi \times \bT, \reali^d)$ and $\MM \in C^{\m-2}(\Gi, \Tuu)$
are the functions uniquely defined by the following equations:
\beq s = \omega {\partial \v \over \partial \te}~, \qquad
\v(I,\te_0) = 0 \quad \mbox{for all $I \in \Gi$;} \label{st} \feq
\beq \q := {\partial \v \over \partial I} f + {\partial \v \over \partial \te} g~;
\label{dq}  \feq
\beq \p = \overline{\p} + \omega {\partial \w \over \partial \te}~, \qquad
\w(I,\te_0) = 0 \quad \mbox{for all $I \in \Gi$}~; \label{qw} \feq
\beq \u := {\partial \w \over \partial I} f + {\partial \w \over \partial \te} g~; \qquad
\MM := {\partial^2 \overline{f} \over \partial I^2}\, \overline{f}  - \left(\FF\right)^2 ~. \label{em} \feq
Furthermore, $\GG \in C^{\m-2}(\Gam, \Tuu)$ and $\HH
\in C^{\m-2}(\Gam,\Tud)$ are two functions such that,
for all $(I, \delta I) \in \Gam$,
\beq \overline{p}(I + \delta I) = \overline{p}(I) + \GG(I, \delta I) \delta I~, \label{equel0} \feq
\beq \overline{f}(I + \delta I) = \overline{f}(I) + \FF(I) \delta I
+ {1 \over 2} \HH(I,\delta I) \delta I^2~,
\quad \HH^{i}_{j k}(I, \delta I) = \HH^{i}_{k j}(I, \delta I)~. \label{equel} \feq
\end{prop}
\textbf{Proof.} See \cite{uno}. \fine
\begin{prop}
\rm{
\textbf{Remark.} In dimension $d=1$, Eqs. \rref{equel0} \rref{equel} can be uniquely solved for
$\GG(I, \delta I)$ and $\HH(I, \delta I)$; in any dimension we have the solutions given by Taylor's formula, i.e.,
\beq \GG(I, \delta I) := \int_{0}^{1} d \var \, {\partial \overline{p}
\over \partial I}(I + \var \delta I)~,
\quad \HH(I, \delta I) := 2 \int_{0}^{1} d \var \, (1 - \var) {\partial^2
\overline{f} \over \partial I^2}(I + \var \delta I)~.
\label{tayf} \feq
If $\overline{p}$ (resp. $\overline{f}$) is a polynomial or rational function of the actions, $\GG$
(resp. $\HH$) can be obtained in a simpler way by direct inspection of Eq. \rref{equel0}
(resp. \rref{equel}). }
\end{prop}
\salto
Now, from the integral equation \rref{inseq} for the function $t \mapsto \L(t)$ we wish to infer
a system of integral inequalities for the functions $t \mapsto | \L(t) |^\mu$,
where $(|~|^\mu)$ is any separating family of seminorms on $\reali^d$. This requires
a set of auxiliary functions, estimating several characters in \rref{inseq}, which are
introduced hereafter.
\salto
\textbf{2C. New auxiliary functions.}
For each set $Z$, we write
\beq Z^{M} := \{ z = (z^\mu)_{\mu \in M}~|~z^{\mu} \in Z~~\forall \, \mu \}~. \feq
For $J \in \reali^d$ and $\varrho = (\varrho^\mu)
\in [0,+\infty]^M$, we put
\beq B(J,\varrho) := \{ I \in \reali^d~|~|I - J|^{\mu} < \varrho^\mu ~~\forall \, \mu \in M \}~. \feq
We further assume the following. \parnn
(i) $\rho = (\rho^{\mu}) \in C([0,U), [0,+\infty]^M)$ is a function such that
\beq B(\J(\tau), \rho(\tau)) \subset \Lambda \qquad \mbox{for
$\tau \in [0,U)$}~. \label{recall} \feq
We put
\beq \Sgr := \{ (\tau, r)~\in [0,U) \times [0,+\infty)^M~|~
r^\mu < \ro^\mu(\tau)~\forall \, \mu \in M \}~. \label{dero} \feq
(ii) $a^\mu, b^\mu,c^\mu,\d^{\mu}_{\nu},\e^{\mu}_{\nu \kap}  \in
C(\Sgr, [0,+\infty))~~ (\mu,\nu,\kap \in M)$
are functions such that for any $\tau \in [0,U)$,  $\delta J \in B(0, \rho(\tau))$ and $\te \in \bT$,
\beq | s(\J(\tau) + \delta J,\te) - \R(\tau) s(I_0,\te_0) - \K(\tau) |^\mu
\leqs a^\mu(\tau, | \delta J|)~, \label{fa} \feq
\beq \big| \w(\J(\tau) +\delta J, \te) - \FF(\J(\tau)) \, \v(\J(\tau) +
\delta J, \te) \big|^\mu~
\leqs b^\mu(\tau, |\delta J|)~, \label{fb} \feq
\beq \big| \u(\J(\tau)+\delta J, \te) -\FF(\J(\tau)) (\w
+\q)(\J(\tau)+\delta J,\te)   \label{fc} \feq
$$ - \MM(\J(\tau)) \v(\J(\tau)+\delta J,\te) \big|^\mu
\leqs c^\mu(\tau,|\delta J|)~,$$
\beq | \GG(\J(\tau),\delta J)|^{\mu}_{\nu} \leqs
\d^{\mu}_{\nu}(\tau,|\delta J|)~, \label{fd} \feq
\beq | \HH(\J(\tau),\delta J)|^{\mu}_{\nu \kap} \leqs \e^{\mu}_{\nu
\kap}(\tau,|\delta J|)~. \label{fe} \feq
In the above, one always intends
\beq | \delta J | := (| \delta J |^{\lambda})_{\lambda \in M}~. \feq
The functions $c^{\mu}, d^{\mu}_{\nu},
e^{\mu}_{\nu \kap}$ are assumed to be nondecreasing with respect to the
variable $r$, i.e.,
\beq (\tau,r), (\tau, r') \in \Sgr,~~~r^{\lambda} \leqs r'^{\lambda}~
\forall \, \lambda \in M ~~~\Rightarrow~~~ c^\mu(\tau,r) \leqs
c^\mu(\tau,r') \label{monot} \feq and similarly for
$d^{\mu}_{\nu}$ and $\e^{\mu}_{\nu\kap}$. Given
$a^{\mu}, ...., \e^{\mu}_{\nu \kap}$, we define the functions
\beq \alpha^{\mu} \in C(\Sgr, [0,+\infty)), \quad
\alpha^\mu (\tau,r) := a^{\mu}(\tau,r)  + \ep b^{\mu}(\tau,r)~,\label{al} \feq
\beq \gamma^{\mu} \in C(\Sgr \times [0,+\infty)^M, [0,+\infty)), \label{ga} \feq
$$ \gamma^{\mu}(\tau,r,\ell) := c^{\mu}(\tau, r) +
\d^{\mu}_{\nu}(\tau, r) \ell^\nu +
{1 \over 2} \e^{\mu}_{\nu \kap}(\tau,r) \ell^\nu \ell^\kap~. $$
In the sequel we will set $\alpha := (\alpha^\mu) \in C(\Sgr, [0,+\infty)^M)$, and intend
$\gamma$ similarly. \parnn
(iii) $R^{\mu}_{\nu}$,  $P^{\mu}_{\nu} \in C([0,U), [0,+\infty))$ ($\mu, \nu \in M$) are functions such that,
for $\tau \in [0,U)$,
\beq | \R(\tau) |^{\mu}_{\nu} \leqs R^{\mu}_{\nu}(\tau)~,~~
| \R^{-1}(\tau) |^{\mu}_{\nu} \leqs P^{\mu}_{\nu}(\tau) ~.
\label{rp} \feq
\parn
\begin{prop}
\label{rema1} \rm{ \textbf{Remarks. (a)} In the main following statements about
$| \L(t) |^\mu$ (Propositions \ref{mainprop}, \ref{proprinc}),
the functions $b^\mu, ..., \e^{\mu}_{\nu \kap}$ will
always be multiplied by the small factor $\ep$. For this reason, in
applications one can determine $b^{\mu},...,e^{\mu}_{\nu \kap}$
via fairly rough majorizations of the left-hand sides of Eqs.
\rref{fb}-\rref{fe}. The situation is different for the functions
$a^{\mu}$, that are not multiplied by $\ep$ and so require
accurate estimates. \parnn
\textbf{(b)} A trivial choice for the
functions in (iii) is $R^{\mu}_{\nu}(\tau) := | \R(\tau)
|^{\mu}_{\nu}$, $P^{\mu}_{\nu}(\tau) := | \R^{-1}(\tau)
|^{\mu}_{\nu}$. This is not satisfactory if one wants more than
the $C^0$ regularity: in fact, this choice does not grant
$R^{\mu}_{\nu}$ and $P^{\mu}_{\nu}$ to be $C^k$ for any $k \geqs 1$. On the other
hand, $C^k$ regularity with $k=1$ or $2$ will be required by some subsequent
manipulations, and in view of this we leave $R^{\mu}_{\nu}$ and
$P^{\mu}_{\nu}$ unspecified. }
\end{prop}
\salto
\textbf{2D. Integral inequalities for $(| \L |^\mu)$.} We keep the assumptions and notations of
the previous paragraph.
\begin{prop}
\label{seclem}
\textbf{Lemma.} Assume that the solution $(\I,\Te)$ of
the perturbed system exists on $[0,U/\ep)$ and that $|\L(t)|^\mu <
\ro^\mu(\ep t)/\ep$
for all $\mu \in M$, $t \in [0,U/\ep)$,
Then, for all $\mu$ and $t$ as above,
\beq | \L(t) |^\mu  \leqs \alpha^\mu(\ep t,\ep | \L(t) |)
+ \ep^2 R^{\mu}_{\lambda}(\ep t) \int_{0}^t d t' P^{\lambda}_{\kap}(\ep t')
\, \gamma^\kap(\ep t', \ep | \L(t') |, | \L(t') |)~, \label{inseqp} \feq
intending $| \L(t) | := (| \L(t) |^{\nu})_{\nu \in M}$.
\end{prop}
\textbf{Proof.} We take the $\mu$-th seminorm of both sides in Eq.
\rref{inseq}. To estimate the right-hand side, we use the consistency inequalities
\rref{cons1} \rref{cons2}, together with the relation
$| \int_{0}^t d t'.. |^\lambda \leqs \int_{0}^t d t' |.. |^\lambda$;
next, we apply the inequalities \rref{fa}--\rref{fe} with
$\delta J = \I(t) - \J(\ep t) = \ep \L(t)$, and the inequalities \rref{rp}.
In this way we obtain
\vfill \eject \noindent
\beq | \L(t) |^\mu \leqs a^\mu(\ep t, \ep | \L(t) |\,) + \ep\, b^\mu( \ep t,
\ep | \L (t)|\,) + \ep^2 R^{\mu}_{\lambda}(\ep t) \int_{0}^t d t' P^{\lambda}_{\kap}(\ep t')
\label{inseqpqp} \feq
$$ \times \Big(  c^{\kap}(\ep t', \ep |
\L(t')|) + \d^{\kap}_{\nu}(\ep t', \ep | \L(t')| ) \, |\L(t')|^{\nu}
+ {1 \over 2} \, \e^{\kap}_{\nu \vsi}(\ep t', \ep | \L(t')|) \,
|\L(t')|^\nu | \L(t') |^{\vsi}\Big)~. $$
Now, the thesis \rref{inseqp} follows from the definitions
\rref{al},\rref{ga} of $\alpha$, $\gamma$. \fine
\salto
\textbf{2E. A general fact on integral inequalities.} This result
is stated without proof, being a simple variation of similar ones appearing in \cite{uno} \cite{Mitr}. \parn
\begin{prop}
\label{lemxy}
\textbf{Lemma.} Let $T \in (0,+\infty]$, $\delta = (\delta^\mu) \in
C([0,T), [0,+\infty]^M)$ and
\beq \XX := \{ (t, \ell) \in [0,T) \times [0,+\infty)^M~|~ \ell^\mu <
\delta^\mu(t)~ \forall \, \mu~\}, \feq
$$ \YY := \{ (t, t', \ell)~|~t \in [0,T),~ t' \in [0,t],~ (t', \ell) \in
\XX~\}~. $$
Consider two functions $\xx = (\xx^\mu) \in C(\XX, [0,+\infty)^M)$ and
$\yy = (\yy^\mu) \in C(\YY, [0,+\infty)^M)$. Let each function $\yy^\mu$ be
nondecreasing
in the last variable: $\yy^\mu(t,t', \ell') \leqs \yy^\mu(t,t', \ell)$ if
$(t,t',\ell), (t, t', \ell') \in \YY$ and
$\ell'^{\nu} \leqs \ell^\nu$ for all $\nu \in M$; furthermore,
let $\la = (\la^\mu),  \mi = (\mi^\mu) \in
C([0,T), [0,+\infty)^M)$ be such that $\mbox{graph}~ \la$, $\mbox{graph}~ \mi$
$\subset \XX$, and
\beq \la^\mu(0) = 0~, \qquad
\la^\mu(t) \leqs \xx^\mu(t, \la(t)) + \int_{0}^t d t' \yy^\mu(t, t',
\la(t'))~, \label{ero} \feq
\beq \mi^\mu(t) > \xx^\mu(t, \mi(t)) + \int_{0}^t d t' \yy^\mu(t, t',
\mi(t'))~ \label{esi} \feq
for all $\mu \in M$, $t \in [0,T)$. Then, for all such $\mu$ and $t$,
\beq \la^\mu(t) < \mi^\mu(t)~.\label{ete} \feq
\end{prop}
\salto
\textbf{2F. The main Proposition.}
We still assume that the solution $\J$ of the averaged system
exists on $[0,U)$, and
define $\R,\K$ via Eqs. \rref{sistr} \rref{sistk}. Moreover, let us be given
a set of functions $\rho^\mu, a^\mu, b^\mu, c^\mu, d^{\mu}_{\nu}, e^{\mu}_{\nu\kap}$
as in paragraph 2C; $\alpha^\mu$ and
$\gamma^\mu$ are defined consequently, as indicated therein.
\begin{prop}
\label{mainprop}
\textbf{Proposition.} Assume there is a
function $\en = (\en^\mu) \in C([0,U),[0,+\infty)^M)$ such that, for all
$\mu \in M$ and $\tau \in [0,U)$,
\beq \en^\mu(\tau) < \ro^\mu(\tau)/\ep~, \feq
\beq \en^\mu(\tau) >
\alpha^\mu(\tau,\ep \en(\tau))
+ \ep R^{\mu}_{\lambda}(\tau) \int_{0}^{\tau} d \tau'
P^{\lambda}_{\nu}(\tau')\,  \gamma^\nu(\tau', \ep \en(\tau'), \en(\tau'))~.
\label{inecont} \feq
Then, the solution $(\I,\Te)$ of the perturbed system exists on
$[0,U/\ep)$; furthermore, defining $\L$ as in Eq. \rref{defr} we have
\beq | \L(t) |^\mu < \en^\mu(\ep t) \qquad \mbox{for all $\mu \in M$, $t
\in [0,U/\ep)$.} \feq
\end{prop}
\textbf{Proof.} It is given in detail in Appendix \ref{apmain}; however,
here we sketch it in few lines. The main idea is to compare the inequalities
\rref{inseqp} for $| \L(t) |^\mu$ and \rref{inecont} for $\en^\mu$, writing
the second one with the change of variables $\tau = \ep t$, $\tau' = \ep t'$.
The thesis follows using Lemma \ref{lemxy} with
$\la^\mu(t) := |\L(t)|^\mu$, $\mi^\mu(t) := \en^\mu(\ep t)$
and obvious choices for $\xx^\mu$, $\yy^\mu$; this is
combined with a continuation principle for ODEs, to
prove the existence of $(\I,\Te)$ for all $t \in [0,U/\ep)$. \fine
\salto
\textbf{2G. A differential reformulation.} We keep
the assumptions at the beginning of the previous paragraph, but we require some more
regularity on the functions
$a^\mu,..., e^\mu_{\nu \kap}$, $P^{\mu}_{\nu}$, $R^{\mu}_{\nu}$  fulfilling Eqs. \rref{fa}-\rref{fe} and
\rref{rp}, namely
$$ a^\mu, b^\mu \in C^2(\Sgr,\reali)~, \qquad
c^\mu, \d^\mu_{\nu}, \e^\mu_{\nu \kap} \in C^1(\Sgr, \reali)~, $$
\beq  R^{\mu}_{\nu} \in C^2([0,U),\reali),
\qquad P^{\mu}_{\nu} \in C^1([0,U), \reali)~. \label{assug} \feq
\begin{prop}
\textbf{Proposition.}
\label{proprinc}
(i) Assume there are $\ellu = (\ellu^\mu) \in [0+\infty)^M$, $(\M^\mu_{\nu}) \in [0,+\infty)^{M^2}$
and $\mm = (\mm^\mu) \in (0+\infty)^M$, such that
\beq \Sigma := \Pi_{\mu \in M} [\ellu^\mu - \mm^\mu, \ellu^\mu + \mm^\mu]
\subset
\Pi_{\mu \in M} (0,\ro^\mu(0)/\ep)~, \label{setsi} \feq
\beq \Lip := \max_{\mu \in M} \sum_{\nu \in M} \M^{\mu}_{\nu} <
1/ \ep~, \label{hj} \feq
\beq \left|{\partial \alpha^\mu \over \partial r^\nu}(0, \ep \ell) \right|
\leqs \M^\mu_\nu
\qquad \mbox{for $\mu, \nu \in M$, $\ell \in \Sigma$}, \label{hi} \feq
\beq | \alpha^\mu(0, \ep \ellu) - \ellu^\mu | + \ep \M^\mu_\nu \mm^\nu <
\mm^\mu
\qquad \mbox{for $\mu \in M$.} \label{hip} \feq
Then, there is a unique $\ell_0 = (\ell^\mu_0)$ such that
\beq \ell_0 \in \Sigma~,
\qquad \alpha(0, \ep \ell_0) = \ell_0~. \label{fixedp} \feq
(ii) With $\ell_0$ as above, let
$\em = (\em^\mu), \en = (\en^\nu) \in C^1([0,U),\reali^M)$ solve the Cauchy
problem
\beq {d \em^\mu \over d \tau} =
P^{\mu}_{\kap} \, \gamma^\kap(\cdot , \ep \en, \en)~, \qquad
\em^\mu(0) = 0~, \label{tre} \feq
$$ {d \en^\mu \over d \tau} = \Big(1 - \ep {\partial \alpha \over
\partial r}\, (\cdot, \ep \en)
\Big)^{-1, \mu}_{~~~ \lambda}
\left( {\partial \alpha^\lambda \over \partial \tau}\,(\cdot , \ep \en)
+ \ep R^{\lambda}_{\nu} P^{\nu}_{\kap}
\, \gamma^\kap(\cdot , \ep \en, \en) + \ep {d R^{\lambda}_{\nu} \over d
\tau} \em^\nu \right)~,
$$
\beq \qquad \en^\mu(0) = \ell^\mu_0  \label{quattro} \feq
($\mu \in M$) with the domain conditions
\beq 0 < \en^\mu < \ro^\mu/\ep~, \qquad
\det \Big(1 - \ep {\partial \alpha \over \partial r}\,(\cdot, \ep \en)
\Big) > 0  \label{due} \feq
(in the above, $1 - \ep {\partial \alpha/\partial r}$ stands for
the matrix $(\delta^{\mu}_{\nu} - \ep {\partial \alpha^\mu /\partial r^\nu})$
($\mu,\nu \in M$), and Eq. \rref{quattro} contains the matrix elements of
its inverse. Note that \rref{tre} implies $\em^\mu \geqs 0$). \parnn
Then, the solution $(\I,\Te)$ of the perturbed system exists on
$[0,U/\ep)$ and
\beq | \L(t) |^\mu \leqs \en^\mu(\ep t) \qquad \mbox{for all $\mu \in M$,
$t \in [0,U/\ep)$.} \feq
\end{prop}
\textbf{Proof.} See Appendix \ref{apprinc}, also containing a preliminary Lemma. \fine
\begin{prop}
\label{rema2}
\rm{
\textbf{Remark.}
The previous Proposition mentions $\ell_0$, the unique fixed point of the map
$\alpha(0, \ep \cdot)$ in the set $\Sigma$ of Eq. \rref{setsi}. The proof in the Appendix
indicates that $\alpha(0, \ep \cdot): \Sigma \vain \Sigma$ has Lipschitz constant $\ep \Lip < 1$
in the maximum component norm $\| z \| := \max_{\mu} | z^\mu |$, with $\Lip$ as in \rref{hj}. So,
from the standard theory of contractions, we have the iterative construction
\beq \ell_0 = \lim_{n \vain +\infty} l_n, \qquad l_1~ \mbox{any point of $\Sigma$}~,~
l_{n} := \alpha(0, \ep l_{n-1}) \qquad \mbox{for $n=2,3,...$}~. \label{iterates} \feq
For each $n \geqs 2$,
\beq \| \ell_0 - l_n \| \leqs (\ep \Lip)^{n-1} {\| l_2 - l_1 \| \over (1 - \ep \Lip)}~. \feq
}
\end{prop}
\salto
\textbf{2H. Implementing the scheme in a typical case: the
"$\EN$-operation".} Let us consider a situation in which (for given data $(I_0, \te_0)$
and $\ep > 0$) we have analytical expressions for the solution $\J$ of the averaged
system \rref{av} (on an interval $[0,U)$) and for all the auxiliary functions
$\R, \K, s, p, ..., \GG, \HH$, $a^\mu,..., e^\mu_{\nu \kap}$, $P^{\mu}_{\nu}$, $R^{\mu}_{\nu}$
of the previous paragraphs (having chosen
a separating family of seminorms $(|~|^\mu)_{\mu \in M}$, and making the regularity assumptions \rref{assug}).
One can provide nontrivial examples where these expressions can be obtained: one of them is considered
in Section \ref{esrigi}. \parn
In this situation, to obtain the final estimates $|\L(t)|^\mu \leqs \en^\mu(\ep t)$ of Proposition
\ref{proprinc} we need: the fixed point $\ell_0$, defined by Eq. \rref{fixedp};
the functions $\em = (\em^\mu), \en = (\en^\mu)$ fulfilling the Cauchy problem \rref{tre} \rref{quattro}.
Typically, to find $\ell_0$ and $\em^\mu, \en^\nu$ analytically will be difficult or impossible,
and a numerical approach will be required. Concerning
$\ell_0$, one can compute numerically the iterates $l_2, l_3, ... l_n$ in
\rref{iterates} up to a sufficiently large order $n$, and then
approximate $\ell_0$ with $l_n$.
As for $\em^\mu, \en^\nu$, one can attack the Cauchy problem
\rref{tre} \rref{quattro} by any package for
the numerical integration of ODEs (paying attention
to the domain conditions \rref{due}). \parn
From now on, the term "$\EN$-operation" will be employed to indicate the numerical
determination of $\ell_0, \em, \en$ along the above lines ({\footnote{this is somehow different from the
"$\Np$-operation" of \cite{uno}, that also included the numerical computation of $\J,
\R, \K$ on $[0,U)$ and was designed to work with a single norm on $\reali^d$.}}).
Generally, the $\EN$-operation to
find $\ell_0$ and $\em, \en$ on the interval $[0,U)$ is faster
(and more reliable) than the computation of $\L(t) := [\I(t) - \J(\ep t)]/\ep$
on the long interval $t \in [0,U/\ep)$, through a direct numerical attack to the perturbed system
\rref{pert}: we think that this gives a practical value to the general framework developed here.
This situation will be exemplified in Section \ref{esrigi}. \salto
\textbf{2I. The "$\EL$-operation".} This expression will be used
to indicate the direct numerical computation of
$\L$ from the perturbed system \rref{pert}, on the time interval $[0,U/\ep)$;
in the general framework of this paper, this operation must be performed only if one wants to test
the efficiency of the $\EN$-operation. \parn
Let us clarify the previous statements, assuming again to have the analytical expressions
of all the functions mentioned in paragraph 2H. Having the expression of $\J$,
we substitute $\I(t) = \J(\ep t) + \ep \L(t)$ in Eqs. \rref{pert}
for $(\I,\Te)$; this gives rise to the Cauchy problem
\beq \left\{\barray{ll} (d \L/ d t) (t) = f(\J(\ep t) + \ep \L(t), \Te(t)) -
\overline{f}(\J(\ep t)), & \quad\L(0) = 0~, \\
(d \Te/ d t)(t) = \om(\J(\ep t) ) +\ep g(\J(\ep t) + \ep \L(t),\Te(t)),
& \quad\Te(0) = \te_0 \farray \right. \label{pertel} \feq
for the unknown functions $t \mapsto (\L(t), \Te(t))$. By definition, the "$\EL$-operation"
is the numerical solution of \rref{pertel} for $t \in [0,U/\ep)$
({\footnote{This differs
from the "$\Lp$-operation" of \cite{uno}, which included a preliminary numerical determination
of $\J$ on $[0,U).$}}). \parn
The efficiency of the $\EN$ operation is tested via $\EL$ comparing: (1) the
CPU times $\TT_{\EL}$, $\TT_{\EN}$ required to perform both operations on
standard machines; (2) the graphs of the functions $| \L^\mu |$ and of their
estimators $\en^\mu$, made available by the two operations. Of course, the test is satisfactory if: \parnn
(i) $\TT_{\EN}$ is considerably shorter than $\TT_{\EL}$; \parnn
(ii) for each $\mu \in M$
the estimator $t \mapsto \en^\mu(\ep t)$ approximates well the envelope
of the rapidly oscillating function $t \mapsto |\L(t)|^\mu$, for $t \in [0,U/\ep)$.
\parn
The whole procedures concerning $\EN$ and $\EL$ are illustrated in the next section;
in the example therein, both (i) and (ii) will occur in the test of $\EN$ via $\EL$.
\vskip 0.4cm \noindent
\vfill \eject \noindent
\section{An example from rigid body dynamics.}
\label{esrigi}
\textbf{3A. Introducing the example.}
We consider a perturbed integrable system of the form \rref{pert}, with
\beq d= 2, \qquad \Lambda := \{ I = (\ha,\ka)~|~\ha, \ka \in (0,+\infty) \},
\qquad \om(I) = \ha \ka, \label{omeq} \feq
$$ f(I,\te) := \big(- \ha (\lau + \mu \cos(2 \te)), - \ka (\lad - \mu \cos(2 \te)\big)~,
\quad g(I,\te) := \mu \sin(2 \te)~; $$
this depends on three real coefficients $\mu,\lau,\lad$ such that
\beq \lau > 0~, \qquad - \lau <  \mu <  \lau~, \qquad \lad > - \lau ~. \label{assxyp} \feq
This system has already appeared in \cite{uno} (Section 4, Example 4) where it
was related to Euler's equation for a rigid body with
gyroscopic symmetry under a damping moment proportional to $\ep$, with a particular dependence on the angular
velocity. The actions $I^1, I^2$
have different physical meaning: in fact, $I^1$ is the equatorial angular velocity (in suitable
units) and $I^2$ measures the inclination of the angular velocity on the gyroscopic axis
({\footnote{See Eq. (4.18) of the cited work.}}).
We will take for \rref{pert} the initial conditions
\beq I_0 = (I^1_0, I^2_0) \in \Lambda~, \qquad \te_0 := 0~. \feq
The forthcoming analysis shows that,
depending on the data and on the other parameters involved in the problem, the numerical values of the solution components
$\I^i$ ($i=1,2$) can be very different over long times; the same happens for the components
$\J^i$ and $\L^i(t) = [\I^i(t) - \J^i(\ep t)]/\ep$. For this reasons, and for the different meaning
of the two actions, it can be of interest to
derive separate estimates for the absolute values $| \L^i(t) |$ ($i=1,2$); this will mark a
difference with the analysis of \cite{uno}, where we only gave a global estimate for
$\sqrt{(L^1)^2 + (\L^2)^2}$. \salto
\textbf{3B. Analysis of the example.}
The average $\overline{f}$ and the solutions $\J$ of \rref{av}, $\R, \K$
of \rref{sistr} \rref{sistk}
(on any interval $[0,U)$) are written in the forthcoming Table 1, which also reports the
auxiliary functions $s,...,\HH$ required by our method.
As anticipated, our aim is to estimate separately
the absolute values $| \L^i |$ ($i=1,2$); this marks
the difference with respect to \cite{uno}, where this example was treated estimating the Euclidean norm
$\sqrt{{(\L^1)}^2 + ({\L^2})^2}$.  In the language of
Section \ref{intro}, analyzing the components of $\L$ corresponds to use the seminorms \rref{casei}, i.e.,
\beq |~ |^i : \reali^2 \vain [0,+\infty)~, \qquad X \mapsto | X |^i := | X^i |
\qquad \qquad (i=1,2)~. \feq
Whenever necessary, we will use for $T^1_1(\reali^2)$,
$T^1_2(\reali^2)$ the consistent seminorms \rref{consof}. \parn
The second half of Table 1 contains
the functions $\rho^i, a^i,..., d^{i}_{j}, e^{i}_{j k}, R^{i}_{j}, P^{i}_{j}$ ($i,j,k=1,2$)
required by the general framework of the previous section.
The choice of $\rho^i$ is an obvious consequence of the form of $\Lambda$;
$a^i, b^i, c^i$ have been computed binding the left-hand sides of Eqs. \rref{fa}-\rref{fc}
by elementary means (similar to the ones employed for the Euclidean norm
estimates of \cite{uno}, but here applied to the components); we observe that $a^i(\tau, r)$
is just a bound on $|s(\J(\tau), \delta J)|^i$ (for $|\delta J|^k = r^k$), because
$\K(\tau)$ and $s(I_0, \te_0)$ in the left-hand side of \rref{fa} are zero.
The functions $d^{i}_{j}, e^{i}_{j k}$ are identically zero
due to the vanishing of $\GG, \HH$; the expressions for $R^{i}_{j}$ and $P^{i}_{j}$
are just the ones of the matrix elements of $\R$ and $\R^{-1}$. \salto
\textbf{3C. Results.} Starting from the functions in Table 1, the $\EN$-operation
has been performed for two choices of the initial data $I^1_0, I^2_0$ and of the parameters
$U$, $\ep$, $\la_1, \la_2, \mu$, producing as a main output the estimators
$\en = (\en^i)_{i=1,2}$ on $[0,U)$; the $\EL$-operation has been performed as a test, producing
directly the function $\L = (L^i)_{i=1,2}$. Both operations were carried over on a PC, using
the MATHEMATICA package. \parn
The results are summarized in Figures a,b,c,d. These report the CPU times $\TT_{\EN}$, $\TT_{\EL}$
(in seconds) and allow to compare the actual values of
$| \L^i(t)|$ with our estimators $\en^i(\tau)$, for $t = \tau/\ep$ and $\tau \in [0,U)$. \parn
It turns out that $\TT_{\EN}/\TT_{\EL} \simeq 1/12$ for the first choice of the parameters
(Figures a, b) and $\TT_{\EN}/\TT_{\EL} \simeq 1/600$ for the second choice
(Figures c, d), where the time scale $U/\ep$ is overwhelmingly long.
In all cases the functions $\tau \mapsto \en^i(\tau)$ practically coincide with the envelopes
of the oscillating functions $\tau \mapsto |\L^i(\tau/\ep)|$, for $\tau \in [0,U)$. \parn
\begin{table}
\textbf{Table 1. A list of functions for the example.} \vskip 0.3cm \hrule
\vskip 0.2cm
For $I = (I^1, I^2) \in (0,+\infty)^2$, $\te \in \bT$ and $\delta I = (\delta I^1,
\delta I^2) \in (-I^1,+\infty) \times (-I^2, +\infty)$ : \vskip 0.2cm \noindent
$ \overline{f}(I) = (-\lau \ha, - \lad \ka)~; $ \parn
$ s(I,\te) =  \dd{\mu \over 2} \sin(2 \te) \left(-{1\over \ka}~, {1 \over \ha}\right),
\qquad v(I,\te) = \dd{\mu \over 2 \ha \ka} \sin^2 \te  \left(-{1\over \ka}, {1 \over \ha}\right)~, $ \parn
$p(I,\te) = \dd{\mu \sin(2 \te) \over 2} \left(- {\lad + \mu \cos(2 \te) \over \ka},
{\lau + 3 \mu \cos(2 \te) \over \ha}\right), \qquad
\overline{p}(I) = (0,0) ~,$ \parn
$ q(I,\te) = \dd{\mu \sin^2 \te \over 2 \ha \ka} \left( - {2 \lad + 2 \mu + \lau + \mu \cos(2 \te) \over \ka},
{\lad + 2 \mu + 2 \lau + 3 \mu \cos(2 \te) \over \ha} \right), $ \parn
$ w(I,\te) = \dd{\mu \sin^2 \te \over 2 \ha \ka} \left( - {\lad + \mu \cos^2 \te \over \ka},
{\lau + 3 \mu \cos^2 \te \over \ha} \right)~, \qquad u(I,\te) = \dd{\mu \sin^2 \te \over 4
\,  \ha \ka}  $ \parn
$ \times \Bigg(
- \dd{4 \lad^2 + 6 \lad \mu + 2 \lad \lau +  \mu \lau +  \mu (4 \lad + 3 \mu + \lau) \cos(2 \te) +
3 \mu^2 \cos^2(2 \te) \over \ka}~, $ \parn
$ \dd{3 \lad \mu + 2 \lad \lau + 10 \mu \lau + 4 \lau^2 + 3 \mu (\lad + 5 \mu + 4 \lau) \cos(2 \te) +
15 \mu^2 \cos^2(2 \te) \over \ha}\Bigg)~; $
\vskip 0.1cm \noindent
$ \FF(I) = \mbox{diag}(- \lau, - \lad)~,~~\MM(I) = \mbox{diag}(- \lau^2, - \lad^2)~,~~
\GG(I, \delta I) = 0,~~\HH(I,\delta I) = 0~. $
\vskip 0.3cm \noindent
For $\tau \in [0,U)$: \vskip 0.1cm \noindent
$\J(\tau) = (I^{1}_0 \, e^{- \lambda_1 \tau}, I^{2}_0 \, e^{- \lambda_2 \tau})~; \quad
\R(\tau)= \mbox{diag}(e^{- \lambda_1 \tau}, e^{- \lambda_2 \tau})~, \quad \K(\tau) = (0,0)~; $
\vskip 0.1cm \noindent
$\rho^i(\tau) := \J^i(\tau)$ ($i=1,2$).
\vskip 0.3cm \noindent
For $\tau \in [0,U)$ and $r = (r^1, r^2) \in [0,\J^1(\tau)) \times [0, \J^2(\tau))$~: \vskip 0.1cm \noindent
$a^1(\tau,r) := \dd{| \mu | \over 2 (\J^2(\tau) - r^2)}~, \qquad
a^2(\tau,r) := \dd{| \mu | \over 2 (\J^1(\tau) - r^1)}~;$
\vskip 0.1cm \noindent
$ b^1(\tau,r) :=
~\dd{\ | \mu | (4 \lau + 4 \lad + |\mu|) \over 8 (\J^1(\tau) - r^1) (\J^2(\tau) - r^2)^2}~,
\qquad b^2(\tau,r) :=
~\dd{\ | \mu | (4 \lau + 4 \lad + 3 |\mu|) \over 8 (\J^1(\tau) - r^1)^2 (\J^2(\tau) - r^2)}~; $
\vskip 0.1cm \noindent
$ c^1(\tau,r) :=
~\dd{\ | \mu | (16 (\lau + \lad)^2 + 16 |\mu| (\lau + \lad) + 3 |\mu|^2)
\over 16 (\J^1(\tau) - r^1) (\J^2(\tau) - r^2)^2}~, $
\vskip 0.1cm \noindent
$ c^2(\tau,r) :=
~\dd{\ | \mu | (16 (\lau + \lad)^2 + 16 |\mu| (\lau + \lad) + 15 |\mu|^2)
\over 16 (\J^1(\tau) - r^1)^2 (\J^2(\tau) - r^2)}~; $
\vskip 0.15cm \noindent
$ d^{i}_{j}(\tau, r) := 0~; \qquad e^{i}_{j k}(\tau,r) := 0~; $
\vskip 0.15cm \noindent
$R^{i}_{j}(\tau) := e^{-\lambda_i \tau} \delta^{i}_{j}~; \qquad
P^{i}_{j}(\tau) := e^{\lambda_i \tau} \delta^{i}_{j} \qquad \qquad \qquad (i,j,k=1,2)~. $
\vskip 0.2cm
\noindent \hrule
\end{table}
\vfill \eject \noindent
\begin{figure}
\parbox{3in}{
\includegraphics[
height=2.0in,
width=2.8in
]%
{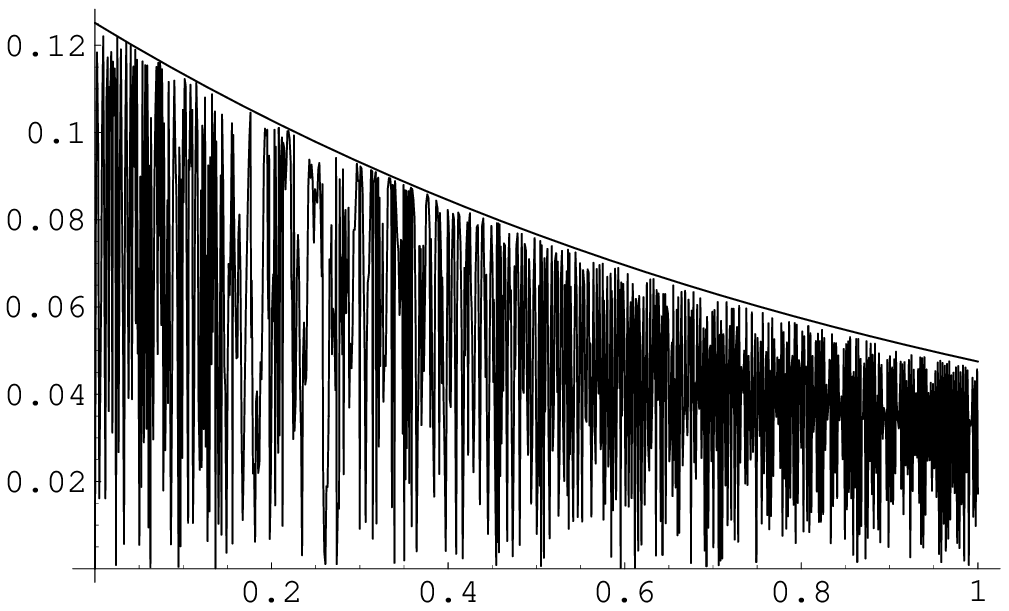}%
\parn
{\textbf{Figure a.} \hbox{$\mu \! = \! 1$}, \hbox{$\lau \! = \! 2$},
\hbox{$\lad \! =\! -1$}, \hbox{$I^1_0 \! =\! 4$}, \hbox{$I^2_0 \! = \! 4$},
\hbox{$\ep \! = \! 10^{-2}$}, \hbox{$U \! = \! 1$}.~ \hbox{$\TT_{\EN} \! = \! 0.031 s$},
\hbox{$\TT_{\EL} \! = \! 0.391 s$. Graphs of $\en^1(\tau)$ and $|\L^1(\tau/\ep)|$}. \parnn}
\label{f1a}
}
\hskip 0.4cm
\parbox{3in}{
\includegraphics[
height=2.0in,
width=2.8in
]%
{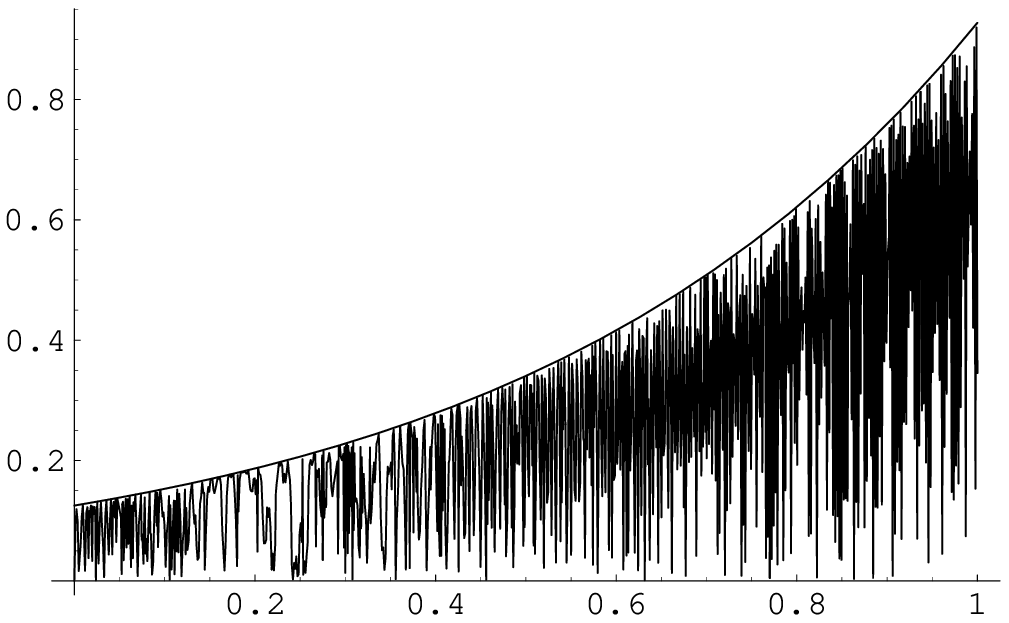}%
\parn
{\textbf{Figure b.} The same parameters as in Fig.a.
Graphs of $\en^2(\tau)$ and $|\L^2(\tau/\ep)|$. \parnn {~} \parnn}
\label{f1d}
}
\parbox{3in}{
\includegraphics[
height=2.0in,
width=2.8in
]%
{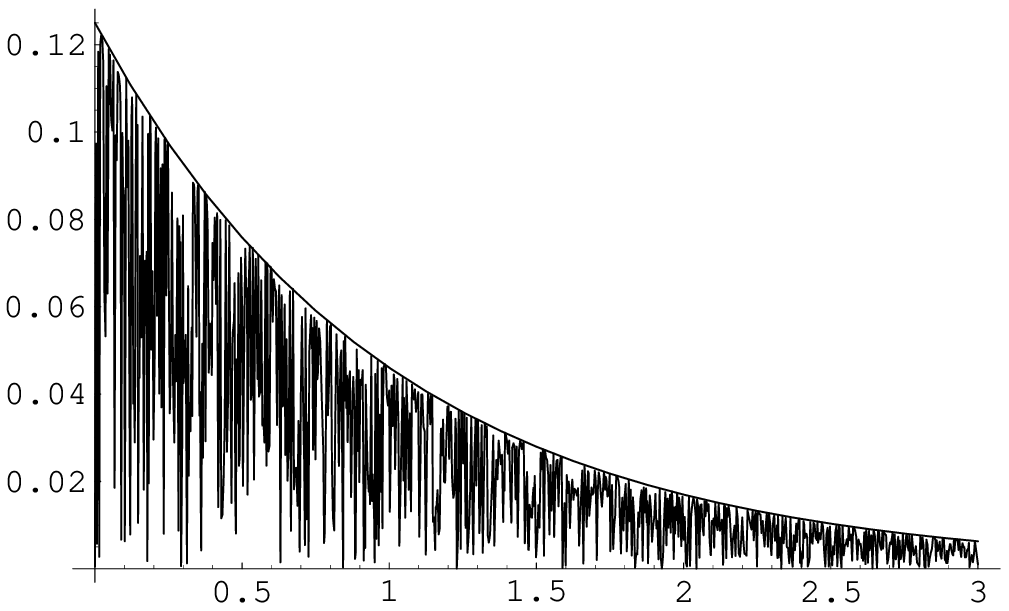}%
\parn
{\textbf{Figure c.} $\mu \! = \! 1$, $\lau \! = \! 1.1$, $\lad \! =\! -1$, $I^1_0 \! =\! 4$, $I^2_0 \! = \! 4$,
$\ep \! = \! 10^{-3}$, $U \! = \! 3$. $\TT_{\EN} \! = \! 0.047 s$, $\TT_{\EL} \! = \! 31.3 s$.
Graphs of $\en^1(\tau)$ and $|\L^1(\tau/\ep)|$. \parnn}
\label{f1b}
}
\hskip 0.4cm
\parbox{3in}{
\includegraphics[
height=2.0in,
width=2.8in
]%
{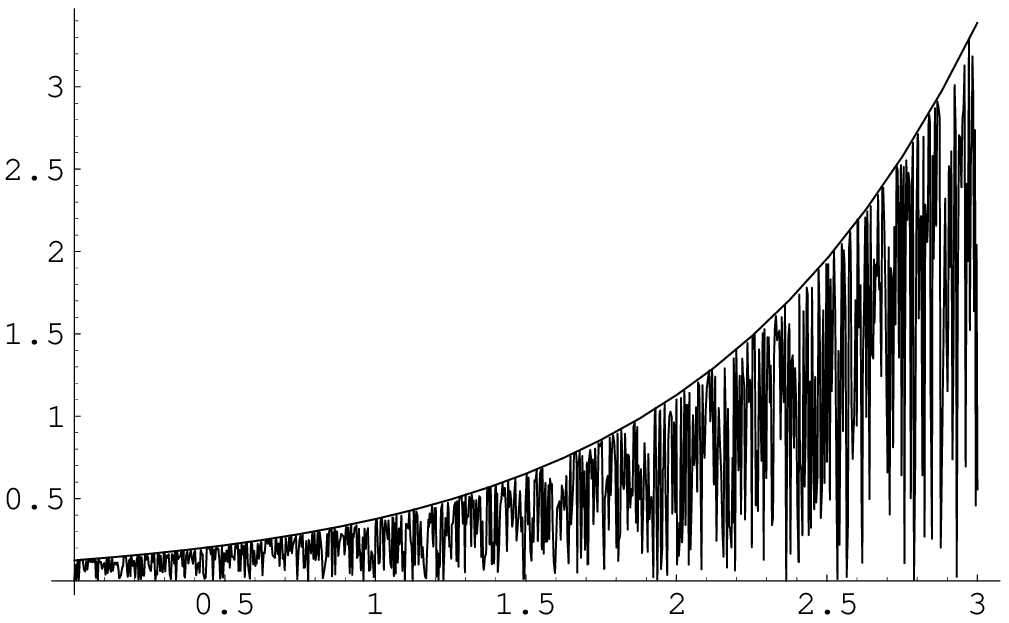}%
\parn
{\textbf{Figure d.} The same parameters as in Fig.c.
Graphs of $\en^2(\tau)$ and $|\L^2(\tau/\ep)|$. \parnn {~} \parnn}
\label{f1e}
}
\end{figure}
\vfill \eject \noindent
\appendix
\section{Appendix. Proof of Proposition \ref{mainprop}.}
\label{apmain}
Let $[0,V/\ep)$ (with $V \in [0,+\infty]$) be the
domain of the maximal solution $(\I,\Te)$ of \rref{pert}, and put
\beq U' := \min(V, U)~. \feq
\textsl{Step 1.  One has $| \L(t) |^\mu < \en^\mu(\ep t)$ for all $\mu \in M$, $t
\in [0,U'/\ep)$}.
To show this, we write the integral inequality \rref{inecont} with $\tau=
\ep t$, $\tau' = \ep t'$; this gives
\beq \en^\mu(\ep t) >
\alpha^\mu(\ep t,\ep \en(\ep t))
+ \ep^2 R^{\mu}_{\lambda}(\ep t) \int_{0}^{\ep t} d t'
P^{\lambda}_{\nu}(\ep t')
\, \gamma^\nu(\ep t', \ep \en(\ep t'), \en(\ep t'))
\label{ineconttt} \feq
for $t \in [0,U/\ep)$, hence for $t \in [0,U'/\ep)$.
\parn
To go on, we use Lemma \ref{seclem} with the constant $U$ therein replaced
by $U'$; this
gives Eq. \rref{inseqp} for $t \in [0,U'/\ep)$. Keeping in mind
Eqs. \rref{inseqp} and \rref{ineconttt},
we apply Lemma \ref{lemxy} with $T := {U' / \ep}$,
$\delta^\mu(t) := \ro^\mu(\ep t)/\ep$, $\xx^\mu(t,\ell) :=
\alpha^\mu(\ep t,\ep \ell)$, $\yy^\mu(t, t', \ell) := \ep^2\, R^{\mu}_{\nu}(\ep t)
P^{\nu}_{\kap}(\ep t') \gamma^\kap(\ep t', \ep \ell, \ell)$,
$\la^\mu(t) := |\L(t)|^\mu$, $\mi^\mu(t) := \en^\mu(\ep t)$
(the requirement $\la^\mu(0) = 0$ is fulfilled by construction). Lemma \ref{lemxy} gives $\la^\mu(t) <
\mi^\mu(t)$, yielding the thesis. \parnn
\textsl{Step 2. One has $V \geqs U$, i.e., $U' = U$
(thus $(\I, \Te)$ exists on $[0,U/\ep)$ and the inequality of
Step 1 holds in this interval)}.
Indeed, suppose $V < U$ and put
\beq K := \{ (t, I) \in [0,V/\ep] \times \reali^d~|~| I - \J(\ep t)|^\mu
\leqs \ep \en^\mu(\ep t) \quad
\forall \, \mu \in M \}~. \feq
This is a compact subset of $\reali \times \reali^d$ and (due to Step 1 and
\rref{recall}) $\mbox{graph}~(\I, \Te) \subset K \times \bT$ $\subset \reali \times \Lambda \times \bT$. But
$K \times \bT$ is compact:
so, by the continuation principle  for ODEs \cite{Zei},
the solution $(\I, \Te)$ can be extended to an interval larger than
$[0,V/\ep)$. This contradicts
our maximality assumption, and concludes the proof. \fine
\vskip 0.4cm \noindent
\section{Appendix. Proof of Proposition \ref{proprinc}.}
\label{apprinc}
We begin with a Lemma, holding under the assumptions at the beginning
of paragraph 2E. Its proof is very similar to the one given in
\cite{uno} for Lemma C.1, so it will not be reported.
\begin{prop}
\label{cored}
\textbf{Lemma.} Assume there are functions
$\en_{\delta} = (\en^\mu_{\delta})
\in C([0,U_{\delta}), [0,+\infty)^M)$, labeled by a parameter $\delta \in
(0,\delta_{*}]$,
such that the following holds: \parnn
(i) $U_{\delta} \vain U$ for $\delta \vain 0^{+}$; \parnn
(ii) for all $\delta \in (0,\delta_*]$, $\mu \in M$  and $\tau \in
[0,U_{\delta})$\,,
\beq \en^\mu_{\delta}(\tau) < \ro^\mu(\tau)/\ep~, \feq
\beq \en^\mu_{\delta}(\tau) = \delta +
\alpha^\mu(\tau,\ep \en_{\delta}(\tau))
+ \ep R^\mu_{\lambda}(\tau) \int_{0}^{\tau} d \tau'
P^{\lambda}_{\kap}(\tau') \gamma^\kap(\tau',
\ep \en_{\delta}(\tau'), \en_{\delta}(\tau'))~;  \label{edecont} \feq
(iii) for each fixed $\tau \in [0,U)$, the limit
\beq \en(\tau) := \lim_{\delta \vain 0^{+}}
\en_{\delta}(\tau) \feq
exists in $[0,+\infty)^M$. \parn
Then the solution $(\I,\Te)$ of \rref{pert} exists
in $[0,U/\ep)$ and, defining $\L$ as in Eq. \rref{defr},
\beq | \L(t) |^\mu \leqs \en^\mu(\ep t) \qquad \mbox{for all $\mu \in M$,
$t \in [0,U/\ep)$.} \feq
\end{prop}
We come to Proposition \ref{proprinc}; so, we make the assumptions at the beginning
of paragraph 2G, strengthened by the smoothness requirements \rref{assug}
for the functions $a^\mu, ..., e^{\mu}_{\nu \kap}$.
\salto
\textbf{Proof of Proposition \ref{proprinc}.} We introduce the norm
$\|~\|$ on $\reali^M$, the function $\aa$ and its generalizations $\aad$
($\delta \in [0,+\infty)$) setting
\beq \| z \| := \max_{\mu \in M} | z^\mu |~; \label{nz} \feq
\beq \aa : \Sigma \vain \reali, \qquad \ell \mapsto \aa(\ell) :=
\alpha(0, \ep \ell)~,  \label{aaa} \feq
\beq \aad : \Sigma \vain \reali~, \qquad \ell \mapsto \aad(\ell) :=
\aa(\ell) + \delta~. \label{aade} \feq
\textsl{Step 1. For each $\delta \geqs 0$, $\aad$ is a contractive map with
respect to the norm $\|~\|$.}
In fact, by \rref{hi} it is
\beq \left| {\partial \alpha^\mu_\delta \over \partial \ell^\nu}(\ell)
\right| =
\ep \left|{\partial \alpha^\mu \over \partial r^\nu}(0, \ep \ell) \right|
\leqs \ep \M^{\mu}_{\nu}
\qquad \mbox{for all $\ell \in \Sigma$}~. \label{a1} \feq
Let $\ell, \ell' \in \Sigma$; the equation
$\alpha^\mu_\delta(\ell) - \alpha^\mu_\delta(\ell') = \int_{0}^1 d s
(\partial \alpha^\mu_\delta /\partial r^\nu)((1 - s) \ell' + s \ell)$
$\times (\ell^\nu - \ell'^\nu)$ implies
\beq | \alpha^{\mu}_{\delta}(\ell) - \alpha^{\mu}_{\delta}(\ell') | \leqs
\ep \M^\mu_\nu | \ell^\nu - \ell'^\nu | ~, \label{eppl} \feq
whence
\beq || \alpha_{\delta}(\ell) - \alpha_{\delta}(\ell') || \leqs
\ep \Lip || \ell- \ell' || <
|| \ell - \ell' ||~. \label{eppll} \feq
The last two inequalities depend on \rref{hj}; contractivity of
$\alpha_\delta$ is proved. \parnn
\textsl{Step 2. There is $\delta_{*} > 0$ such that,
for all $\delta \in [0, \delta_{*}]$, $\aad$ sends $\Sigma$ into itself.}
In fact, for any $\delta \geqs 0$ and $\ell \in \Sigma$,
$$ | \alpha^\mu_\delta(\ell) - \ellu^\mu | =
| \alpha^\mu_0(\ell) + \delta - \ellu^\mu | \leqs
| \alpha^\mu_0(\ell) - \alpha^\mu_0(\ellu) | + | \alpha^\mu_0(\ellu) -
\ellu^\mu |
+ \delta  $$
\beq \leqs \ep \M^\mu_{\nu} | \ell^\nu - \ellu^\nu | +
| \alpha^\mu_0(\ellu) - \ellu^\mu | +
\delta \leqs \ep \M^\mu_\nu \mm^\nu + | \alpha^\mu_0(\ellu) - \ellu^\mu |
+ \delta~,
\label{ea} \feq
where the second inequality follows from Eq. \rref{eppl} with $\delta=0$.
To go on, we note that \rref{hip} implies the existence of a $\delta_{*} >0$
such that
\beq | \alpha^\mu(0, \ep \ellu) - \ellu^\mu | + \ep \M^\mu_\nu \mm^\nu +
\delta_{*} \leqs \mm^\mu~
\qquad \mbox{for all $\mu \in M$.} \label{hipp} \feq
For $\delta \in [0,\delta_{*}]$ and $\ell \in \Sigma$, Eqs.
\rref{ea}, \rref{hipp} imply $|\alpha^\mu_\delta(\ell) - \ellu^\mu | \leqs
\mm^\mu$,
i.e., $\aad(\ell) \in \Sigma$. \parnn
\textsl{Step 3. For all $\delta \in [0, \delta_{*}]$, the map $\aad$ has a
unique
fixed point $\ell_{\delta} = (\ell^\mu_\delta) \in \Sigma$, which depends
continuously on $\delta$}. Existence
and uniqueness of the fixed point follows from the Banach theorem on
contractions;
to prove continuity we note that, for all $\delta, \delta' \in [0,
\delta_{*}]$,
\beq ||\ell_{\delta} - \ell_{\delta'}|| = || \aad(\ell_{\delta}) -
\aadp(\ell_{\delta'}) || =
|| \aa(\ell_{\delta}) + \delta - \aa(\ell_{\delta'}) - \delta' ||  \feq
$$ \leqs || \aa(\ell_{\delta}) - \aa(\ell_{\delta'}) || +
| \delta - \delta' | \leqs \ep \Lip || \ell_{\delta} -
\ell_{\delta'} ||
+ | \delta - \delta' |~, $$
the last inequality depending on \rref{eppll} with $\delta=0$. This implies
\beq || \ell_{\delta} - \ell_{\delta'}|| \leqs {| \delta - \delta' | \over
1 - \ep \Lip}~; \feq
so the map $\delta \mapsto \ell_{\delta}$ is Lipschitz, and a fortiori
continuous. \parnn
\textsl{Step 4. Proving the thesis  of (i).} This follows from Step 3, with
$\delta=0$. \parnn
\textsl{Step 5. Proving the thesis of (ii)}. For any $\delta \in [0,
\delta_{*}]$, let $\ell_{\delta}$ be as in
Step 3.
First of all, let us consider the space of real matrices $N =
(N^\mu_\nu)_{\mu, \nu \in M}$, with the norm
$\| N \| := \sup_{z \in \reali^M, z \neq 0} \| N z \|/\| z \|$, and note
that the second
inequality \rref{a1}, with \rref{hj}, yields
\beq \ep \| {\partial \alpha \over \partial r}(0, \ell_\delta) \| \leqs \ep
\Lip < 1~. \feq
By a well-known fact on matrices of the form $1 - N$, this implies
\beq \det (1 - \ep {\partial \alpha \over \partial r}(0, \ell_\delta)) > 0~. \label{toric} \feq
From the standard continuity theorems for
the solutions of a parameter-dependent Cauchy problem, we know
that there is a family $(U_{\delta}, \em_{\delta},
\en_{\delta})_{\delta \in (0,\delta_{*}]}$ with the forthcoming
properties (a) (b):  \parnn (a) for all $\delta \in (0,\delta_{*}]$,
it is $\em_{\delta} = (\em^\mu_{\delta}),
\en_{\delta} = (\en^\mu_{\delta}) \in
C^1([0,U_{\delta}),\reali^M)$; furthermore,
\beq {d \em^\mu_\delta \over d \tau} =
P^{\mu}_{\kap} \gamma^\kap(\cdot, \ep \en_\delta, \en_\delta)~, \qquad
\em^\mu_\delta(0) = 0~, \label{tred} \feq
$$ {d \en^\mu_\delta \over d \tau} =
\Big(1 - \ep {\partial \alpha \over \partial r}\, (\cdot, \ep \en_\delta
)\Big)^{-1, \mu}_{~~~\lambda}
\left({\partial \alpha^\lambda \over \partial \tau}\,(\cdot , \ep \en_\delta )
+ \ep R^{\lambda}_{\nu} P^{\nu}_\kap \, \gamma^\kap(\cdot , \ep
\en_\delta , \en_\delta ) +
\ep {d R^\lambda_\nu \over d \tau} \em^\nu_\delta \right), $$
\beq \qquad \en^\mu_\delta (0) = \ell^\mu_\delta \label{quattrod} \feq
\beq 0 < \en^\mu_{\delta} < \ro^\mu/\ep~, \qquad
\det (1 - \ep {\partial \alpha \over \partial r}(\cdot, \ep
\en_\delta)) >0  \label{dued} \feq
(concerning the domain conditions in the last line, recall \rref{toric};
$\em^\mu_\delta \geqs 0$ by \rref{tred}). \parnn
(b) One has
\beq U_{\delta} \vain_{\delta \vain 0^{+}} U, ~~\en^\mu_\delta(t)
\vain_{\delta \vain 0^{+}} \en^\mu(t),~~ \em^\mu_\delta(t) \vain_{\delta
\vain 0^{+}} \em^\mu(t)~~
\mbox{for all $t \in [0, U)$}, \label{du} \feq
where $U$, $\em^\mu$, $\en^\mu$ are as stated in (ii). \parn
Let us consider the pair $\em_{\delta}, \en_{\delta}$ for any $\delta \in
(0, \delta_{*}]$. Then, integrating
\rref{tred},
\beq \em^\mu_{\delta}(\tau) = \int_{0}^{\tau} d \tau' P^{\mu}_{\nu}(\tau')
\gamma^\nu(\tau', \ep \en_\delta(\tau'), \en_\delta(\tau')) \qquad
\mbox{for $\tau \in [0, U_{\delta})$.}
\label{lass} \feq
Furthermore, from Eq. \rref{quattrod} and \rref{tred} we infer
$$ 0 = \Big(1 - \ep {\partial \alpha \over \partial r}(\cdot,
\ep \en_\delta)\Big)^{\mu}_{\vsi} {d \en^\vsi_\delta \over d \tau} -
\left({\partial \alpha^\mu \over \partial \tau}\,(\cdot , \ep \en_\delta )
+ \ep R^{\mu}_{\nu} P^{\nu}_{\kap} \, \gamma^\kap(\cdot , \ep
\en_\delta , \en_\delta ) +
\ep {d R^{\mu}_{\nu} \over d \tau}~\em^\nu_\delta \right)  $$
$$ = {d \en_\delta^\mu \over d \tau} -
\ep {\partial \alpha^\mu \over \partial r^\vsi}(\cdot, \ep
\en_\delta)  {d \en^\vsi_\delta \over d \tau} -
\left({\partial \alpha^\mu \over \partial \tau}\,(\cdot , \ep \en_\delta )
+ \ep R^{\mu}_{\nu} {d \em^\nu_\delta \over d \tau} +
\ep {d R^{\mu}_{\nu} \over d \tau}~\em^\nu_\delta \right) $$
\beq = {d \over d \tau} ( \en^\mu_{\delta} - \alpha^\mu(\cdot, \ep
\en_{\delta}) -
\ep R^\mu_\nu \em^\nu_{\delta} )~; \label{quattrodd} \feq
therefore, for $\tau \in [0,U)$,
$$ \en^\mu_{\delta}(\tau) - \alpha^\mu(\tau, \ep \en_{\delta}(\tau)) -
\ep R^\mu_\nu(\tau) \, \em^\nu_\delta(\tau)
= \en^\mu_{\delta}(0) - \alpha^\mu(0, \ep \en_{\delta}(0)) -
\ep R^\mu_\nu(0) \, \em^\nu_\delta(0) $$
\beq = \ell^\mu_{\delta} - \alpha^\mu(0, \ep \ell_\delta) = \ell^\mu_{\delta} -
\alpha^\mu_0(\ell_{\delta}) = \delta \label{las} \feq
(recall the initial conditions in Eqs. \rref{tred} \rref{quattrod}, Eqs. \rref{aaa}
\rref{aade} and Step 3, giving
$\ell^\mu_\delta = \alpha^\mu_\delta(\ell_\delta) =
\alpha^\mu_0(\ell_\delta) + \delta$). \parn
From Eqs. \rref{las} \rref{lass} we see that $\en_\delta$ fulfils Eq.
\rref{edecont} of Lemma \ref{cored}. Due to Eq. \rref{du} on
the limit for $\delta \vain 0^{+}$, from the cited Lemma we
obtain the thesis. \fine \salto
\vskip 0.4cm \noindent
\textbf{Acknowledgments.} This work has been partially supported by the GNFM
of Istituto Nazionale di Alta Ma\-te\-ma\-ti\-ca and by MIUR,
Research Project Cofin/2006 "Metodi geometrici nella teoria delle onde non
lineari e applicazioni".
\vfill \eject 
\end{document}